\documentstyle[12pt]{article}                                        
\begin{document}
\thispagestyle{empty}
\begin{center}
{\large\bf Negative-energy perturbations \\
 in cylindrical equilibria \vspace{1mm}\\
 with a radial electric field} \vspace{6mm}\\
{\large G. N. Throumoulopoulos\footnote
{Permanent  affiliation: Section of Theoretical Physics,
 Physics Department, University of Ioannina 
 GR  451 10 Ioannina, Greece} and D. Pfirsch \vspace{1mm}\\
   Max-Planck-Institut f\"{u}r Plasmaphysik, EURATOM
Association \\ \vspace{1mm}
 D-85748 Garching, Germany \\ \vspace{6mm}
July 1997}  
\end{center}

\vspace{12cm}
\noindent
To be published in Physical Review E
\newpage
\begin{center}
{\large\bf Abstract} 
\end{center}
\setcounter{page}{1}

The impact of an equilibrium radial electric field $\bf E $
on  negative-energy perturbations (NEPs) in 
cylindrical equilibria
of magnetically confined plasmas is investigated within the
framework of Maxwell-drift kinetic theory. It turns out that
for wave vectors  with a non-vanishing
component parallel to the magnetic field 
the conditions for the 
existence of NEPs in equilibria with $\bf E=0$
[G. N. Throumoulopoulos and D. Pfirsch, Phys. Rev. E {\bf 53}, 2767 (1996)] 
remain valid,  while the condition for the existence of perpendicular NEPs, 
which are found to be the most important perturbations, 
is  modified.
  For 
 $|e_i\phi|\approx T_i$, a scaling which is satisfied in the
edge region of magnetic confinement systems ($\phi$ is the electrostatic
potential), 
 the impact of $\bf E$ on perpendicular NEPs depends on the
 value  of $T_i/T_e$,  i.e.,
 a) for $T_i/T_e < \beta_c\approx P/(B^2/8\pi)$ ($P$ is the total plasma
pressure) the electric field 
 does not have any effect and b)
for $T_i/T_e> \beta_c$, a case which is of operational
 interest in magnetic confinement
systems,   the existence of perpendicular NEPs depends on
$e_\nu {\bf E}$, where $e_\nu$ is 
the charge  of the particle species $\nu$.
In the latter case, for tokamaklike equilibria
and  H mode parameters pertaining to the  plasma edge two
regimes of  NEPs exist. 
In the one of them  the critical value  $2/3$
of $\eta_i\equiv 
\partial \ln T_i/\partial\ln N_i$
plays a role in the existence of ion NEPs,  
as in equilibria with  ${\bf E}={\bf 0}$,
while  a critical value of $\eta_e$ does 
not occur  for the existence
 of electron NEPs.
However, $\bf E$ has  a ``stabilizing" effect on  both particle species 
in that (a) the portion
of   particles associated with NEPs (active 
particles) is nearly independent of the plasma magnetic properties,
i.e.,  it is nearly the same in a diamagnetic plasma and 
in a paramagnetic plasma,
while in equilibria with ${\bf E}={\bf 0}$
this portion is much larger in a paramagnetic plasma than
 in a diamagnetic plasma, and (b)  the fraction of active   
 particles can decrease from the plasma interior to the edge,  
e.g., for the case of electron NEPs in an equilibrium of a diamagnetic 
plasma, 
contrary to
equilibria with ${\bf E}={\bf 0} $. 
In particular, 
the 
fraction  of active electrons
decreases with increasing $\eta_e$ and for $\eta_e>\eta_0\approx 4/3$ 
the electric field stabilizes the  electrons,  
in that the fraction of active electrons
becomes  smaller than the one corresponding to equilibria with
${\bf E}={\bf 0}$. 
In addition,  
$\bf E$ has  similar stabilizing
effects  on electron NEPs in stellaratorlike  equilibria
with  pressure profiles identical to those of tokamaklike-equilibria, 
while it results
in an increase of the fraction of active ions in 
 reversed-field-pinchlike equilibria.
The present results indicate that the radial electric field 
reduces the NEPs
activity in the edge region of tokamaks and stellarators, 
the reduction of electron NEPs
being more pronounced than that of ion NEPs.
\newpage
\begin{center} 
{\bf I.\ \ INTRODUCTION }
\end{center}
Negative-energy waves  are potentially dangerous because  
they  can 
lead to either  linear instability 
\cite{BrMo}  
or  nonlinear, explosive instability 
\cite{We} - \cite {Un}. 
Expressions for the second variation of the free energy 
$F^{(2)}$ were derived by Pfirsch and Morrison  \cite{PfMo}
for arbitrary perturbations of general equilibria 
within the framework of 
  dissipationless   Maxwell-Vlasov 
and drift kinetic theories. It was also also found that
 negative-energy perturbations 
exist in any Maxwell-Vlasov equilibrium whenever the unperturbed distribution
 function $f_\nu^{(0)}$ of any particle species $\nu$ deviates
from monotonicity and/or isotropy in the vicinity 
of a single point, i. e., whenever the condition 
\begin{equation}
\left({\bf k}\cdot{\bf v}\right)
\left( {\bf k}\cdot 
\frac{\partial f_\nu^{(0)}}{\partial {\bf v}}\right) >0 
                                                     \label{-1}
\end{equation}
holds (in the frame of reference of minimum equilibrium energy)
for any particle species $\nu$ for some position vector ${\bf x}$
and velocity $\bf v$ and for some local vector ${\bf k}$.
The proof of this result 
was based on infinitely strongly localized
perturbations, which correspond to $|{\bf k}|\rightarrow \infty$.
This raises the question of the degree of localization actually required
for NEPs to exist in a certain equilibrium. Studying a 
homogeneous  Maxwell-Vlasov plasma  \cite{CoPf92},
force-free equilibria  with a sheared magnetic filed \cite{CoPf93}
and general one- and two-dimensional equilibria of magnetically
confined plasmas \cite{CoPf94a}-\cite{CoPf97},   
Correa-Restrepo and Pfirsch showed that NEPs
exist for any deviation of the equilibrium distribution function of
any of the species from monotonicity and/or isotropy,
without having to impose any restricting conditions on $\bf k$.

NEPs which are  not strongly localized can be investigated
more conveniently in the framework of Maxwell-drift kinetic theory
which  eliminates from the outset all perturbations
with perpendicular wavelengths smaller than  the gyro-radius.
In the context of this theory, for a homogeneous magnetized plasma
 it was found 
that  NEPs exist for
any wave vector ${\bf k}$ with a non-vanishing component parallel to
the magnetic field (parallel and oblique modes) whenever the condition
\begin{equation} 
v_{\parallel} \frac{\partial f^{(0)}_{g\nu}}{\partial v_{\parallel}}
>0                                                        
						     \label{0}
\end{equation} 
is satisfied  for the equilibrium guiding center distribution
function $f^{(0)}_{g\nu}$
for some particle species $\nu$ and parallel velocity $v_{\parallel}$ in the
frame of lowest equilibrium energy \cite{PfMo}.  
For  the more interesting cases of inhomogeneous magnetically confined
plasmas and equilibria depending on just one Cartesian coordinate $x$ 
\cite{ThPf94} and 
cylindrical equilibria 
with vanishing electric fields  
\cite{ThPf96a,ThPf96b} 
in addition to parallel and
oblique modes for which condition (1) also applies, perpendicular NEPs  
are possible. The latter are the most important perturbations because 
they can exist even if 
$v_{\parallel} 
(\partial f^{(0)}_{g\nu}/\partial v_{\parallel})< 0$, 
which  is satisfied e.g.  for
 Maxwellian distribution functions
for all $v_\parallel$. In plane geometry
the pertinent condition is 
$$
\frac{\textstyle dP^{(0)}}{\textstyle d x}   
\frac{\textstyle \partial  f^{(0)}_{g\nu}}{\textstyle\partial x}
<0,
$$ where $P^{(0)}$ is the equilibrium plasma pressure. 
For  tokamaklike  equilibria with  singly peaked pressure profiles   
the existence
of both  ion and electron perpendicular NEPs is associated with 
the critical value $2/3$ of 
the quantity $\eta_{\nu} \equiv \partial \ln T_{\nu}/ \partial \ln N_{\nu}$
($T_{\nu}$ is the temperature and $N_{\nu}$ the density of  particle
species $\nu$) which usually governs the onset of the temperature gradient
driven modes.  For cylindrical equilibria an additional regime of NEPs
exists, related to the curvature of the poloidal magnetic
field.
Also, for the case of  cold-ion   equilibria ($T_i=0$) 
a large portion of  
electrons is associated with NEPs (active particles). 

The purpose of the present paper is two-fold:
(a) to investigate   the impact of a radial electric field
   on NEPs in cylindrical equilibria of magnetically confined plasmas
and  (b) to  extend the study to  equilibria with $T_i\neq 0$. 
The method of investigation
consists in evaluating 
the general expression $F^{(2)}$ for the 
second-order perturbation energy  within the framework 
of the linearized dissipationless Maxwell-drift kinetic theory. 
This is the subject of Sec. II. 
The conditions for the existence of NEPs
are obtained in Sec. III. It turns out that for parallel and oblique 
perturbations condition (\ref{0}) remains valid,
 while the condition for perpendicular
NEPs
 (which remain the  most important perturbations)
is modified.
 To apply the  condition
for perpendicular NEPs in equilibria of magnetic confinement
systems the equilibrium equations are needed,  which are   derived 
 in Sec. IV. 
Shearless stellaratorlike equilibria are possible with local  
Maxwellian distribution functions,
while  tokamaklike and reversed-field-pinchlike 
equilibria  can be obtained from 
shifted Maxwellian
distribution functions, which imply net toroidal currents.
For these kinds of distribution functions
 and H  mode parameters pertaining
to the plasma  edge the condition for the
existence of perpendicular NEPs is applied  
in Sec. V, and 
 the effect of  $\bf E$ on the threshold value
of $\eta_\nu$ is examined.
In Sec. VI the impact of $\bf E$ on the fraction of active particles  
 is investigated   for
 shearless stellaratorlike,   tokamaklike 
and reversed-field pinchlike analytic  
equilibria.
Our conclusions are summarized  in  Sec. VII.
\begin{center}
{\bf II.\ \ EQUILIBRIUM AND SECOND-ORDER PERTURBATION ENERGY}
\end{center}

We start with   a brief outline of the Maxwell-drift kinetic theory  
adapted to the needs of the  present study. More details
can be found in Ref. \cite{PfMo}.

The expression for the free energy $F^{(2)}$  upon arbitrary
perturbations of general equilibria is   given by
\begin{equation}
F^{(2)} = \int d^3 x \ T_0^{(2)0},           \label{P0}
\end{equation}
where  $T_0^{(2)0}$ is the energy component of the second-order
energy-momentum  tensor $T_\rho^{(2)\lambda}$. 
To derive the tensor $T_\rho^{(2)\lambda}$,
Pfirsch and Morrison \cite{PfMo} used the following  modified
Hamilton-Jacobi formalism.
 Let $H_\nu(p_i,q^i,t)$ be
the Hamiltonian for particles of species
$\nu$ for the perturbed state in a phase space $p_1,
...,p_4$, $q^1,...,q^4$ where $(q^1,q^2,q^3)$ are generalized coordinates
so that ${\bf x}={\bf x}(q^1,q^2,q^3)$ and correspondingly 
 ${\bf p}={\bf p}(p_1,p_2,p_3)$, where ${\bf x}$ 
is  the  position  vector in normal space;
 $p_4, q^4$ is an additional pair of canonical variables which
is needed to describe guiding center motion.
Let $H_\nu ^{(0)}(P_i,Q^i)$ be the equilibrium Hamiltonian
 in the phase space $P_1,\ldots, P_4$,
$Q^1,...Q^4,$ and let
$S_\nu(P_i,q^i,t)$ be a mixed-variable generating
function for a canonical
transformation between $p_i$, $q^i$ and $P_i,Q^i$.
The ${\bf x}$, $t$ dependence of $H_\nu$ is given via 
 electromagnetic potentials $\phi({\bf x},t)$ and ${\bf A}({\bf x},
t)$, the electric and
magnetic fields ${\bf E}({\bf x},t)$ and ${\bf B}({\bf x}  ,t)$ and 
their derivatives.
The quantities
$p_i$ and $Q^i$ are obtained from $S_\nu$ as
\begin{equation}
p_i=\frac{\partial S_\nu}{\partial q^i},\ \ Q^i= \frac{\partial
S_\nu}{\partial P_i},                                \label{p.1}
\end{equation}
and $S_\nu$ must be the solution of the modified Hamilton-Jacobi
equation
\begin{equation}
 \frac{\partial S_\nu}{\partial t} + H_\nu
\left(\frac{\partial S_\nu}{\partial q^i},q^i,t\right)=H_\nu^{(0)}
\left(P_i,\frac{\partial S_\nu}{\partial P_i}\right).
                                                    \label{p.2}
\end{equation}
The time-independent, zeroth-order solution $S_\nu^{(0)}$ of Eq.
(\ref{p.2}),
 needed to obtain
 $ T_\rho^{(2)\mu}$,
is then simply given by the identity transformation
$S_\nu^{(0)}=\sum_{\nu}P_iq^i$.

The theory can be derived from the Lagrangian
\begin{eqnarray}
 L&=&-\sum_\nu\int dqdP\varphi_\nu(P_i, q^i, t) 
      \left\lbrack \frac{\partial S_\nu}{\partial t}
       +   H_\nu
\left(\frac{\partial S_\nu}{\partial q^i}, q^i, t \right)
 - H_\nu^{(0)}\left(P_i,\frac{\partial  S_\nu}{\partial P_i}\right)
      \right \rbrack  \nonumber \\                           
 & & -\frac{1}{8\pi}\int d^3x ( {\bf E}^2 - {\bf B}^2 ).
                                                 \label{p.24}
\end{eqnarray}
 Here,
 $dq\, dP\equiv dq^1\cdots dq^4\, dP_1\cdots dP_4$;
 $\varphi_\nu$ are density functions related to the {\em particle} 
 distribution functions $f_\nu$ and the latter are related
to the guiding center distribution functions $f_{g\nu}$ by  
 Eq. (\ref{R 32}) below.
The energy momentum tensor can be obtained by
 using the Euler-Lagrange equations resulting from the variational
principle
\begin{equation}
\delta\int_{t_1}^{t_2} L dt=0,                             \label{p.6}
\end{equation}
(with $\varphi_\nu$, $S_\nu$, $\phi$
and  ${\bf A}$  
the quantities to be varied) and Noether's theorem. 
In the context of the linearized theory one obtains
\begin{eqnarray}
T_\rho^{(2)\lambda}&=& -\sum_\nu\int d\hat{\tilde{q}} d\tilde{P}\left(
    \frac{\partial S_\nu^{(1)}}{\partial\tilde{q}^\rho}-\frac{e_\nu}{c}A_\rho
    ^{(1)}\right)\left[ f_\nu^{(0)}\left(\frac{\partial S_\nu^{(1)}}
    {\partial\tilde{q}^\kappa}-\frac{e_\nu}{c}A_\kappa^{(1)}\right)
    \frac{\partial^2{\cal H}_\nu^{(0)}}
    {\partial \tilde{P}_\lambda\partial\tilde{P}_\kappa}
    \right.
    \nonumber \\ 
& & \left.+f_\nu^{(0)}F_{\tau\sigma}^{(1)}
     \frac{\partial^2{\cal H}_\nu^{(0)}}
    {\partial\tilde{P}_\lambda\partial F_{\tau
     \sigma}^{(0)}}
     +\left(f_\nu^{(0)}
     \frac{\partial S_\nu^{(1)}}{\partial \tilde{P}_i}\right)_{\ ,i}
     \frac{\partial {\cal H}_\nu^{(0)}}
	  {\partial\tilde{P}_\lambda}\right] \nonumber \\ 
& &  -2F_{\mu\rho}^{(1)}\sum_\nu\int d\hat{\tilde{q}} d\tilde{P}\left[
     f_\nu^{(0)}\left(\frac{\partial S_\nu^{(1)}}{\partial\tilde{q}^\kappa}
    -\frac{\textstyle e_\nu}{\textstyle  c} A_\kappa^{(1)}\right)
     \frac{\partial^2{\cal H}_\nu^{(0)}}
     {\partial\tilde{P}_\kappa\partial F^{(0)}
     _{\mu\lambda}}\right. \nonumber \\ 
& &  \left.+f_\nu^{(0)} F_{\sigma\tau}^{(1)}
    \frac{\partial^2{\cal H}_\nu^{(0)}}
    {\partial F_{\mu\lambda}^{(0)}\partial F_{\sigma\tau}
    ^{(0)}}\right]
    -\frac{1}{4\pi} F_{\mu\rho}^{(1)} F^{(1)\mu\lambda}
    \nonumber \\
& &
     +\delta_\rho^\lambda
   \left(\sum_\nu\int d\hat{\tilde{q}} d 
   \tilde{P} f_\nu^{(0)}({\cal H}_\nu^{(2)}
   -{\cal H}_\nu^{(0)(2)}) + \frac{1}{16\pi} F_{\tau\sigma}^{(1)}
    F^{(1)\tau\sigma}\right).
						      \label{P1}
\end{eqnarray}
Here,  
 the superscripts $(0)$, $(1)$ and $(2)$ denote equilibrium, 
first-  and second-order 
quantities;
 the tilde  signifies that the time is included, i.e.
$$(\tilde q^i)= (\tilde q^0,\ldots,\tilde q^4)= (ct,{\bf x}, q^4),$$
$$(\tilde{p}_i)=(\tilde p_0, \ldots,\tilde p_4) = (p_0, {\bf p}, p_4),
\ \ \ \ cp_0=\frac{\partial S_\nu}{\partial t},$$
$$(\tilde Q^i)= (\tilde Q^0,\ldots,\tilde Q^4)= (ct,  {\bf x}, Q^4),$$
$$(\tilde{P}_i)=(\tilde P_0, \ldots,\tilde P_4) = (P_0, {\bf p}, p_4),\ \ \ \
P_0=\mbox{constant},$$ 
$${\cal H}_\nu(\tilde p_i, \tilde q^i)= cp_0 +
 H_\nu(p_1,\ldots, p_4, q^1, \ldots, q^4),$$
$${\cal H}_\nu^{(0)}(\tilde P_i, \tilde Q^i)= cP_0 +
 H_\nu(P_1, \ldots, P_4, Q^1, \ldots,  Q^4);$$
$d\hat{\tilde{q}}\,d\tilde P=dq^4\,dP_1\cdots dP_4$; 
 $A_\mu =\left(-\phi,{\bf A}\right)$ with $A_4\equiv 0$;
 $F_{\mu\nu}$ is the
electromagnetic tensor;
the symbol $C^i_{\ ,j}$ signifies covariant derivative of the
vector ${\bf C}$ with contravariant components $C^i$:
$$
C^i_{\ ,j}\equiv \frac{\partial C^i}{\partial q^j} -  {\Gamma}^i_{jl} C^l
$$
where ${\Gamma}^i_{jl}$ are the Christoffel symbols;  
the scalar quantity
 $\left(f_\nu^{(0)}
     \frac{\textstyle  \partial S_\nu^{(1)}}{\textstyle 
	  \partial \tilde{P}_i}\right)_{ ,i}$,
which replaces
 $\frac{\textstyle \partial}{\textstyle\partial \tilde q_i}\left(f_\nu^{(0)}
     \frac{\textstyle  \partial S_\nu^{(1)}}{\textstyle
          \partial \tilde{P}_i}\right)$ in Eq. (46) of Ref. \cite{PfMo},
 results from the  contraction in the tensor 
$\left(f_\nu^{(0)}
     \frac{\textstyle  \partial S_\nu^{(1)}}
	  {\textstyle  \partial \tilde{P}_i}\right)_{,j}$.

The Hamiltonians 
$ H_\nu$
for the guiding center motion of a particle species $\nu$, which  
appear in  $F^{(2)}$,  are
obtained from  Littlejohn's Lagrangian formulation
of the guiding center theory \cite{Li} in the form 
given by Wimmel \cite{Wim}:
\begin{equation} 
L_\nu=\left(\frac{e_\nu}{c}\right){\bf A}_\nu^{\star}
\cdot\dot{{\bf x}} - e_\nu\phi_\nu^\star 
						\label{P3}
\end{equation} 
with
\begin{displaymath} 
{\bf A}^{\star}_{\nu} = {\bf A} + \frac{m_{\nu}c}{e_\nu}q^4
 {\bf b},
\end{displaymath} 
\begin{displaymath} 
e_{\nu} \phi^{\star}_{\nu} = e_\nu\phi + \mu B +
\frac{m_{\nu}}{2} \left((q^4)^2 +{\bf v}_E^2 \right),
\end{displaymath} 
\begin{displaymath} 
{\bf  v}_{E} = c \frac{{\bf E}
\times{\bf B}}{B^2}, 
\end{displaymath} 
\begin{displaymath} 
{\bf E}=-{\bf \nabla}\phi - \frac{1}{c}\frac{\partial{\bf A}}{\partial t},
\ \ \ \ {\bf B }= \nabla\times {\bf A},\ \  \ \ {\bf b} = \frac{{\bf B}}{B}.
\end{displaymath} 
The Euler-Lagrange equations
yield 
$q^4={\bf v}\cdot{\bf b}=v_\parallel$ 
and the  guiding center velocity
$\dot{\bf x}={\bf v}\equiv {\bf v}_g$  and  $\dot q^4$ as functions
of $t$, ${\bf x}$  and $q^4$:
\begin{equation}
\dot{{\bf x}}={\bf v}=
	      {\bf v}_{g\nu}\left(t,{\bf x},q^4\right) 
	     = \frac{q^4}{ B^\star_{\nu\parallel}}
	      {\bf B}_\nu^\star + \frac{c}{B^{\star}_{\nu\parallel}}
	      {\bf E}^\star_\nu\times{\bf b}
					     \label{R 24}
\end{equation}  
and
\begin{equation} 
 \dot{q}^4=V^4 
		 \left(t,{\bf x},q^4\right) =
	   \frac{e_\nu}{m_\nu}
	   \frac{1}{B^\star_{\nu\parallel}}
	   {\bf E}_\nu^\star\cdot
	   {\bf B_\nu}^\star.
					     \label{R 25}
\end{equation} 
Here, ${\bf E}^\star_\nu \equiv - {\bf \nabla} \phi_\nu^\star - 
\frac{\textstyle   1}{\textstyle   c}
	\frac{\textstyle  \partial{\bf A}^\star_\nu}
	      {\textstyle  \partial t},  \ \ \ 
{\bf B_\nu}^\star\equiv {\bf \nabla} \times{\bf A}^\star_\nu  \ \ 
	 \mbox{and} \ \ 
B^{\star}_{\nu\parallel} \equiv {\bf B}^\star_{\nu}\cdot {\bf b}$.
 The momenta canonically conjugated
to ${\bf x}$ and $q^4$ follow from (\ref{P3}): 
\begin{equation}
{\bf p}=\frac{\partial L_\nu}{\partial \dot{{\bf x}}}
       =\frac{e_\nu}{c}{\bf A}^{\star}_\nu,   \ \  \ 
p_4 = \frac{\partial L_\nu}{\partial\dot{q}^4}= 0.
						       \label{P4}
\end{equation}
Since Eqs. (\ref{P4}) do not contain
$\dot{{\bf x}}$ and  $\dot{q}^4$, they are constraints between the momenta 
and the coordinates. It therefore follows  that
Hamilton's equations based on the usual  Hamiltonians corresponding to
the above non-standard Lagrangians
are not the equations of motion. To overcome this difficulty,
Dirac's  theory of constrained dynamics
\cite{Di}
 is applied,  which yields the Dirac Hamiltonians
\begin{equation}
 H_\nu=e_\nu\phi_\nu^\star + {\bf v_{g\nu}}\cdot
       \left({\bf p}-(e_\nu/c){\bf A}^\star_\nu\right)+V^4p_4.
							    \label{P5}
\end{equation}
Particular solutions of the equations of motion following
from the Hamiltonians (\ref{P5}) are  the
constraints (\ref{P4}). The distribution functions
    $f_\nu({\bf x}, q^4, {\bf p},
p_4, t)$ must guarantee that these constraints are satisfied. 
As concerns this requirement, 
it is important to note that 
${\bf p}-\left(\frac{\textstyle   e_\nu}{\textstyle  
c}\right){\bf A}^\star_\nu=0$ and $p_4=0$ do not represent special values
 of some
constants of motion. Therefore, $\delta$-functions of the constraints are not
constants of motion either. But $f_\nu$ must
be proportional to
such $\delta$-functions and, at the same time,
  also  a constant of motion. Both
conditions are uniquely satisfied by
\begin{equation}
f_\nu=\delta(p_4)\delta
      \left({\bf p}-\frac{e_\nu}{c}{\bf A}^\star_\nu\right)
       B_{\nu\parallel}^\star
       f_{g\nu}\left({\bf x},q^4,\mu,t\right),
					 \label{R 32}
\end{equation}
 where the
guiding center distribution functions $f_{g\nu}$ 
are constants of motion and solutions of the drift
kinetic differential equations 
\begin{equation}
 \frac{\partial f_{g\nu}}{\partial t}+{\bf v}_{g\nu}\cdot
 \frac{\partial f_{g\nu}}{\partial{\bf x}}
	+V^4\frac{\partial f_{g\nu}}{\partial q^4}= 0.
\end{equation}
 
In the present paper 
cylindrical equilibria are considered.  
With the coordinates $q^1, q^2, q^3 $  specified
to be  the  cylindrical 
coordinates 
 $r$, $\theta$, $z$ with unit basis vectors ${\bf e}_r$, 
${\bf e}_\theta$,  ${\bf e}_z$
 the equilibrium  vector potential and
 magnetic field are given by 
\begin{equation}  
{\bf A}^{(0)} =A^{{(0)} }_{\theta}(r){\bf e}_{\theta}+A^{{(0)} }_{z}(r)
{\bf e}_{z}                                                     
						      \label{1}
\end{equation} 
and
\begin{equation} 
{\bf B}^{{(0)} } = B^{{(0)} }_{\theta}(r)
{\bf e}_{\theta} + B^{{(0)} }_{z}(r)
{\bf e}_{z },                                                    
						       \label{2}
\end{equation} 
with
\begin{equation}  
\frac{1}{r} (rA^{{(0)} }_{\theta})^\prime = B^{{(0)} }_{z}, 
\ \ \  (A^{{(0)} }_{z})^{\prime} = -B^{{(0)} }_{\theta},
							\label{2a}
\end{equation} 
where the  prime denotes differentiation with respect to $r$. 
The equilibrium electric field can be expressed in
terms of the scalar potential $\phi^{(0)}(r)$ as
\begin{equation}
{\bf E}^{(0)}= - \nabla \phi^{(0)} =
	       - (\phi^{(0)})^\prime{\bf e}_r.   
						      \label{2b} 
\end{equation}
For the equilibria defined above,  the  guiding center 
velocity  [Eq. (\ref{R 24})] becomes 
\begin{eqnarray}  
{\bf  v}^{{(0)} }_{g\nu}& =& v_\parallel {\bf b}^{{(0)} } 
      + {\bf v}_\perp^{(0)} \nonumber \\  
      & = & v_\parallel {\bf b}^{{(0)} }
      - \frac{c}{e_\nu B^{\star{(0)} }_{\nu\parallel}}
      \left[ e_\nu(\phi^{(0)})^\prime + \mu (B^{(0)})^\prime 
    -\frac{e_\nu v_\parallel B^{\star{(0)} }_{\nu\perp}}{c}
       + \frac{m_\nu}{2} (v_E^{(0)\ 2})^\prime\right]
      \left({\bf e}_r \times {\bf b}^{{(0)} }\right),   \nonumber\\
							    \label{3}
\end{eqnarray} 
where ${\bf v}_\perp^{(0)}={{\bf b}^{(0)} }\times
\left({\bf  v}^{{(0)} }_{g\nu}
\times  {\bf b}^{(0)}\right)$ is the perpendicular component
of ${\bf  v}^{{(0)} }_{g\nu}$ consisting of the 
${\bf E}\times{\bf B}$,  $\nabla B$,  curvature,    and
polarization drifts;  
$B^{\star{(0)} }_{\nu\parallel}\equiv{\bf B}^{\star{(0)} }_{\nu} 
\cdot {\bf b}^{(0)}$ and $B^{\star{(0)} }_{\nu\perp}\equiv
{\bf b}^{(0)}\times\left({\bf B}^{\star{(0)}}_{\nu}\times 
{\bf b}^{(0)}\right)$.  
For thermal particles it holds that $B^{\star{(0)} }_{\nu\parallel}\approx B^{(0)}$
and $\frac{\textstyle B^{\star{(0)}}_{\nu\perp}  }{\textstyle 
B^{\star{(0)}}_{\nu\parallel} } \approx {\cal O} (r_{g\nu}/r_0)$
where
 $r_{g\nu}$  is the thermal Larmor radius for the particle species
$\nu$ and $r_0$ the macroscopic scale length.
 ${\bf v}_{g\nu}^{(0)}$
has no $r$-component and therefore $r$ is a constant of motion.
Since there is also no force parallel to ${\bf B}^{(0)}$, another 
constant of motion is the parallel guiding center velocity
$v_\parallel$. The guiding center distribution functions $f_{g\nu}^{(0)}$
are therefore functions of $r$, $v_\parallel$ and the magnetic moment $\mu$.
From Eq. (\ref{R 25}) it follows that $V_4^{(0)}=0$ and  hence the Dirac 
Hamiltonians [Eq. (\ref{P5})] are written in the form
\begin{equation}
 H_\nu^{(0)}=e_\nu\phi_\nu^{\star{(0)}} + {\bf v_{g\nu}}^{(0)}\cdot
       \left({\bf p}-(e_\nu/c){\bf A}^{\star(0)}_\nu\right).
							    \label{3b}
\end{equation}

The general expression for the second order perturbation energy 
[Eq.(\ref{P0})] 
is evaluated for these   equilibria and for initial perturbations
${\bf A}^{(1)}={\bf 0}$ and $\dot{\bf A}^{(1)}={\bf 0}$. It is also shown
{\em a posteriori} that one can choose initial perturbations  such that
the charge density $\rho^{(1)}$ vanishes 
without changing the particle contributions to the energy. Thus, choosing
perturbations of this kind, we can set from the outset
$F_{\mu\lambda}^{(1)}=A_\rho^{(1)}=0$

After a lengthy
 derivation, which can be conducted along the lines of that
for cylindrical equilibria with $\bf E = 0$ reported in 
detail in Appendix A of Ref. \cite{ThPf96b},
$F^{(2)}$ is cast in the concise
form:
\begin{eqnarray} 
F^{(2)}& = &- \sum_{\nu}\int S(r)dr dv_\parallel d{\mu}
\left\{\frac{B^{\star{(0)} }_{\nu\parallel}}{m_\nu} \left|G^{(1)}_{\nu}\right|^2
\left({\bf k}\cdot{\bf v}^{{(0)} }_{g\nu}\right) \right. 
						      \nonumber \\ 
& &\left.\times \left[ 
\left({\bf b}^{\star (0)}\cdot {\bf k}\right)
\frac{\partial f^{{(0)} }_{g\nu}}{\partial v_\parallel} 
 - \frac{k_\perp }{\omega^{\star{(0)} }_\nu}
\frac{\partial f^{{(0)} }_{g\nu}}{\partial r}\right]\right\}.
						    \label{4}
\end{eqnarray} 
Here,
$\omega_\nu^{\star(0)}\equiv 
(e_\nu B^{\star(0)}_{\nu\parallel})/ (cm_\nu)$,
 $G_\nu^{(1)}(r, q^4, \mu)$ are arbitrary first-order quantities
 relating to the generating functions $S_\nu^{(1)}$
 for the perturbations; 
 ${\bf b}_\nu^{\star(0)}\equiv {\bf B}_\nu^{\star(0)}/
 B_{\nu\parallel}^{\star(0)}$ ;
${\bf k}=k_\theta {\bf e}_\theta
+k_z{\bf e}_z$ 
is the wave vector lying in magnetic surfaces;
$k_\parallel$  and $k_\perp$ are its components
 parallel and perpendicular to ${\bf B}^{(0)}$, respectively, and
$$
S(r)\equiv r\int_{\theta_0}^{\theta_0 + \frac{2\pi }{r k_\theta}}
            \int_{z_0}^{z_0 + \frac{2\pi}{k_z} } d\theta\,d z  
$$
is a normalization surface,
 where  $\theta_0$ and $z_0$ are constants. 
 We note that $F^{(2)}$
depends on $G^{(1)}_{\nu}$ only via $|G^{(1)}_{\nu} |^2$.
The first-order charge density $\rho^{(1)}$ is a $v_\parallel, {\mu}$
integral over an expression that is linear in 
$G^{(1)}_{\nu}$. One can therefore satisfy
the relation ${\rho}^{(1)}=0$
by a proper distribution of positive
and negative values of $G^{(1)}_{\nu}$ on which $F^{(2)}$ does not
depend. 

Compared with the corresponding expression for 
equilibria with ${\bf E}^{(0)}={\bf 0}$ (Eq. (37) of Ref. \cite{ThPf96a}), 
$F^{(2)}$ contains new terms stemming  from 
${\bf b}_\nu^{\star(0)}$ and from the 
${\bf E}\times{\bf B}$ and polarization drift components of
${\bf v}_{g\nu}$. In particular, as  shown in Section III 
the ${\bf E}\times{\bf B}$ drift  modifies the condition for the
existence of NEPs with wave vectors perpendicular to ${\bf B}^{(0)}$.
\begin{center}  
{\bf III. \ \ CONDITIONS FOR THE EXISTENCE
 OF NEGATIVE-ENERGY PERTURBATIONS}
\end{center}
  
 The derivations in this section 
are very similar to those concerning equilibria
 with $\bf E= 0$ \cite{ThPf94,ThPf96a,
ThPf96b} so that  details   need not be  given here.

The  following conditions
  must  be satisfied only locally in
$r$, $v_\parallel$ and $\mu$, and refer to  the frame of
reference of minimum energy.

\noindent
{\em Parallel perturbations ($k_\perp =0$)} \ \ \  
NEPs exist when 
\begin{equation}
  v_\parallel  
\frac{\partial f^{{(0)} }_{g\nu}}{\partial  v_\parallel}>0
						  \label{7}
\end{equation}  
is satisfied  for  at least  one particle species.     
This condition, 
which   was first derived by Pfirsch and Morrison  for a
homogeneous magnetized plasma \cite{PfMo}, 
agrees with those obtained by Correa-Restrepo and Pfirsch for 
several Maxwell-Vlasov equilibria \cite{CoPf92}-\cite{CoPf97}.
\newline

\noindent
{\em Oblique perturbations} ($ k_\parallel \neq 0$ and 
$k_\perp \neq 0$)\ \ \ 
If condition (\ref{7})  is satisfied for at least  one particle species
 $\nu$,
only perturbations with wave vectors satisfying in addition
 the relations
\begin{equation} 
\frac{k_\parallel}{k_\perp}<\min(\Lambda_\nu,\ M_\nu)\ \ \ \mbox{or}\ \
\ \frac{k_\parallel}{k_\perp}>\max(\Lambda_\nu,\  M_\nu),
					     \label{8} 
\end{equation} 
with
\begin{displaymath} 
\Lambda_{\nu} \equiv -\frac{{\bf v}_{g\nu\perp}^{(0)}}{v_\parallel}
\cdot \left({\bf e}_r \times {\bf b}^{{(0)} }\right)
\end{displaymath} 
and
\begin{displaymath} 
M_{\nu} \equiv 
\left(
 \frac{1}{\omega^{\star{(0)} }_{\nu}}
\frac{\partial f^{(0)} _{g\nu}}{\partial r}
-\frac{B_{\nu\perp}^{(\star(0)}}{B_{\nu\parallel}^{(\star(0)}}
\frac{\partial f^{(0)} _{g\nu}}{\partial v_\parallel}\right)
\left(\frac{\partial f^{(0)} _{g\nu}}{\partial v_\parallel}\right)^{-1}
\end{displaymath} 
can have negative energy.
The orders  of magnitude of
$\Lambda_\nu$ and $M_\nu$ depend on the
particle energy.  For example, if
\begin{equation}  
m_\nu v_\parallel^2\approx\mu B^{(0)}\approx |e_\nu \phi|\approx T_\nu
                                                   \label{8a}
\end{equation}
with $\phi(\infty)=0$ it holds that
\begin{equation} 
 |\Lambda_\nu| \approx |M_{\nu}|
\approx \frac{r_{g\nu}}{r_0} <<1.
                                               \label{8b}
\end{equation} 
Relation (\ref{8b}) indicates  that condition 
(\ref{8}) imposes no essential restriction
on the magnitude or the orientation of the wave vectors associated with
 NEPs.

  If 
\begin{equation} 
  v_\parallel   
\frac{\partial f^{{(0)} }_{g\nu}}{\partial  v_\parallel} <0,
					      \label{9}
\end{equation} 
a condition which is satisfied at all points of 
a Maxwellian distribution function,
NEPs exist if, in addition to (\ref{9}), it holds that
\begin{equation} 
\min(\Lambda_\nu,M_\nu) <\frac{k_\parallel}{k_\perp}<\max(\Lambda_\nu,\
M_\nu).
						  \label{10}
\end{equation} 
 For the scaling (\ref{8a}) 
  condition (\ref{10})
 implies that
\begin{equation} 
\frac{k_\parallel}{k_\perp} \approx \frac{r_{g\nu}}{r_0} <<1,
						  \label{11}
\end{equation} 
which indicates that  the most  important NEPs,
in the sense that the less restrictive condition (\ref{9}) is
involved, are associated with  nearly perpendicular wave vectors.
It may be noted that for a homogeneous magnetized plasma in 
thermal equilibrium, although
condition  (\ref{9}) is satisfied, NEPs are not possible because 
 $M_\nu =\Lambda_\nu=0$ and therefore condition (\ref{10})
 is not satisfied. This also follows
from condition (\ref{7}) which is pertinent for the existence of NEPs in 
a homogeneous magnetized plasma and is not satisfied in thermal
equilibrium, i.e., 
for Maxwellian distribution functions.
\newline

\noindent
{\em Perpendicular perturbations} ($k_\parallel=0$) \ \ \ 
In this case 
the second-order perturbation energy  [Eq. (\ref{4})]
reduces to
\begin{equation} 
F^{(2)}=4\pi \sum_\nu\int drd v_\parallel d\mu S(r)|G^{(1)}_\nu|^2
\frac{B^{\star{(0)} }_{\nu\parallel}}{m^2_\nu} 
\frac{W_{\nu\perp}}{\left(B^{(0)} \right)^2}
\left(\frac{k_\perp}{\omega^{\star{(0)} }_\nu}\right)^2 Z_\nu
Q_\nu
					  \label{11a}
\end{equation} 
with
\begin{equation} 
Z_\nu=\frac{(B^{(0)})^2}{4\pi W_{\nu\perp}}
\left[\frac{e_\nu}{c}v_\parallel B^{\star{(0)} }_{\nu\perp}
-e_\nu\phi^\prime -\mu(B^{(0)})^\prime
-\frac{m_\nu}{2}\left(v_E^{(0)\ 2}\right)^\prime\right]
					    \label{12}
\end{equation} 
and
\begin{equation} 
Q_\nu= 
\frac{\partial f^{{(0)} }_{g\nu}}{\partial r} -
\omega_\nu^{\star(0)}
\frac{B^{\star{(0)} }_{\nu\perp}}{B^{\star{(0)} }_{\nu\parallel}}
\frac{\partial f^{{(0)} }_{g\nu}}{\partial  v_\parallel},
					    \label{13}
\end{equation} 
where 
$W_{\nu\perp}=\mu B^{(0)}$ is  
the perpendicular particle energy.
Eq. (\ref{11a}) implies that $F^{(2)} <0$ {\em for any $k_\perp$}
  whenever the condition
\begin{equation}
Z_\nu Q_\nu<0 
					      \label{13b}                   
\end{equation}                                 
is satisfied {\em irrespective of the sign of
 $v_\parallel(\partial f_{g\nu}^{(0)}/\partial v_\parallel)$}.
Therefore, there are two regimes of NEPs
which are determined
by the relations
 \begin{equation} 
Z_\nu<0\ \ \mbox{and}\ \ Q_\nu >0
					    \label{14}
\end{equation} 
and 
\begin{equation} 
Z_\nu>0\ \ \mbox{and}\ \ Q_\nu <0.
					     \label{15}
\end{equation} 
For the evaluation of  conditions (\ref{14}) and (\ref{15})  
the  equilibrium equations 
are required, which      
are constructed in the following section.

\begin{center}  
{\bf IV. \ \ QUASINEUTRAL MAXWELL-DRIFT KINETIC EQUILIBRIUM EQUATIONS}
\end{center}

The equilibria must satisfy 
\begin{equation} 
\nabla\cdot{\bf E}^{(0)} =4\pi \rho^{(0)}
					    \label{16}
\end{equation} 
and
\begin{equation} 
\nabla\times {\bf B}^{(0)} =
		       \frac{4\pi}{c} {\bf j}^{(0)},
					    \label{16a}
\end{equation} 
where the charge density $\rho^{(0)} $ and  current density
 ${\bf j}^{(0)}$
are expressed self-consistently in terms of  the guiding
center distributions functions $f_{g\nu}^{(0)}$
 in the context 
of the Maxwell-drift kinetic theory
 (see Eqs. (8.14) and (8.15)
of Ref. \cite{CoPfWi}). 
For the system under consideration,
owing to the presence of ${\bf E}^{(0)}$, the set of equilibrium 
equations following from Eqs. (\ref{16}) and (\ref{16a}) 
are  rather  complicated. 
For this reason we employ 
 the quasineutral Maxwell-drift kinetic theory  
which can be derived self-consistently  by  dropping the 
electric-field-energy term in the  Lagrangian  (\ref{p.24}).
A similar method was employed in Refs. 
\cite{CoPf93},  \cite{Un} and \cite{PfCo}. 
Consequently, Eq. (\ref{16}) is replaced by the 
quasineutrality condition, which is explicitly  given by 
\begin{equation} 
\sum_\nu e_\nu \int dv_\parallel d \mu B^{\star(0)}_{\nu\parallel}
f_{g\nu}^{(0)}
+ \sum_\nu div\, \int \, dv_\parallel d \mu B^{\star{(0)}}_{\nu\parallel}
f_{g\nu}^{(0)}
\frac{m_\nu c}{B^{(0)}}({\bf v}_{g\nu}^{(0)}-{\bf v}_E^{(0)})
\times {\bf b}^{(0)}=0
						\label{18}
\end{equation} 
and  Eq. (\ref{16a}) by
\begin{eqnarray}
 \sum_\nu e_\nu \int dv_\parallel d \mu B^{\star{(0)}}_{\nu\parallel}
f_{g\nu}^{(0)} {\bf v}_{g\nu}^{(0)}  & &  \nonumber \\ 
 -  \sum_\nu c\  curl\, \int dv_\parallel d \mu B^{\star{(0)}}_{\nu\parallel}
f_{g\nu}^{(0)} 
  \left\lbrack\mu {\bf b}^{(0)} -
 \frac{m_\nu}{B^{(0)}}v_\parallel\left({\bf v}_{g\nu\perp}^{(0)}
  - {\bf v}_E^{(0)}\right) \right. & &
  \nonumber \\
 \left.-\frac{m_\nu c}{(B^{(0)})^2}({\bf v}_{g\nu}^{(0)}
 -{\bf v}_E^{(0)})\times{\bf E}^{(0)}
      +  \frac{2 m_\nu}{B^{(0)}}
\left\{\left({\bf v}_{g\nu}^{(0)}
    -{\bf v}_E^{(0)}\right)\cdot{\bf v}_E^{(0)}\right\}
      {\bf b}^{(0)}\right \rbrack  & & \nonumber \\
      =\frac{c}{4\pi}\nabla\times {\bf B}^{(0)} & &. 
                                                        \label{18a}
\end{eqnarray}
The first terms on the left-hand sides of Eqs. (\ref{18}) and (\ref{18a})
 represent
 guiding-center charge and current density contributions,  respectively,  
 and the other terms
 polarization and magnetization  contributions.

We  consider  equilibria  of the following kind:
\begin{enumerate}
\item The distribution functions are specified to be local shifted Maxwellians
\begin{equation} 
f_{g\nu}^{(0)}=\left(\frac{m_\nu}{2\pi}\right)^{1/2}
\frac{N^{(0)}_\nu (r)}{T^{(0)}_\nu(r)^{3/2}}
  \exp\left\{- \frac{\mu B^{(0)}(r)    +1/2m_\nu
\left[v_\parallel-V_{\nu}^{(0)}(r)\right]^2}{T_\nu^{(0)}(r)}\right\},
						   \label{19}
\end{equation} 
where $N_\nu^{(0)}$ and $T_\nu^{(0)}$ are, respectively, the density and
temperature  for particles of species $\nu$. 
They can describe   
cylindrical  tokamaklike,  reversed-field pinchlike  
and for $V_\nu^{(0)}\equiv 0$ shearless stellaratorlike
plasmas, which are close to thermal equilibrium.
 For the former equilibria the shift velocities
$V_\nu^{(0)}$  satisfy
\begin{equation} 
\frac{V_\nu^{(0)}}{(v_\nu)_{th}} \approx
\frac{r_{g\nu}}{r_0}\ll 1
				    \label{20}
\end{equation} 
and, as shown later, lead to a non-vanishing toroidal current. 

\item Since a radial electric field may play a role in the L-H
transition of
 magnetic  confinement systems, e.g. \cite{GrBu,FiFu},
we adopt for the ion
electrostatic energy the scaling
\begin{equation} 
|e_i \phi^{(0)}| \approx T_i^{(0)}  \ \ \ \ \phi^{(0)}(\infty)=0,
				    \label{22}
\end{equation}
which is satisfied in the edge region.  
\end{enumerate}
Using the above assumptions,
neglecting  small terms of  the order $r_{g\nu}/r_0$ 
and suppressing the superscript $(0)$ from the equilibrium
quantities,
Eq. (\ref{18}) and the $\theta$- and $z$- components of Eq. (\ref{18a}),
respectively,  
yield 
\begin{equation} 
\sum_\nu e_\nu N_\nu = 0,
				    \label{24}
\end{equation} 
\begin{equation} 
j_\theta =b_\theta  e_i N_i (V_i-V_e) + c\frac{b_z}{B}
	  P^\prime = -\frac{c}{4\pi} B_z^\prime
		                 		    \label{25}
\end{equation} 
and
\begin{equation} 
j_z =b_z e_i N_i (V_i-V_e) - c\frac{b_\theta}{B}
	  P^\prime = -\frac{c}{4\pi}\frac{1}{r}(rB_\theta)^\prime,
				    \label{26}
\end{equation} 
where 
\begin{equation} 
P\equiv\sum \int dv_\parallel d\mu B^\star_{\nu \parallel}\mu B f_{g\nu}
 =\sum_\nu N_\nu T_\nu
                                                      \label{25a} 
\end{equation}
For $V_\nu\equiv 0$ for all $\nu$ Eqs. (\ref{25}) and (\ref{26}),
 respectively, reduce 
to 
\begin{equation} 
\frac{b_z}{B}P^\prime =-\frac{B_z^\prime}{4\pi} 
				    \label{27}
\end{equation} 
and
\begin{equation} 
 \frac{b_\theta}{B}P^\prime =-\frac{1}{4\pi}\frac{1}{r}(r B_\theta)^\prime.
				    \label{28}
\end{equation} 
The solutions of  Eqs. 
(\ref{27}) and (\ref{28}) satisfy the relation 
$B_\theta=a\frac{\textstyle B_z}{\textstyle r}$,
with $a= \mbox{constant}$.  They are singular at $r=0$ and therefore  $a=0$. 
For $B_\theta\equiv 0$ Eq. (\ref{28}) is 
satisfied identically and the only possible equilibrium,
which is  described by Eq. (\ref{27}), 
is a $\theta$-pinch or shearless stellaratorlike configuration
with vanishing axial current, a case which is examined in Sec. VI.

Multiplying Eqs. (\ref{25}) and (\ref{26}) by the integrating
factors $B_z$  and $B_\theta$, respectively, and adding 
the resulting equations, one obtains  the pressure balance
relation
\begin{equation} 
\frac{d}{dr}\left( P + \frac{B^2}{8\pi}\right) 
+ \frac{B_\theta^2}{4\pi r}=0,                                    
					  \label{29}
\end{equation} 
which will be used in place of Eq. (\ref{26}).
 
Summarizing, quasineutral equilibria can be  described by the set of Eqs.   
(\ref{24}), (\ref{25}), (\ref{25a}) and (\ref{29}). 
Four out of the eight functions involved must be assigned, e.g.,
$P(r)$, $B_z(r)$, the shift velocity difference $V_i(r)-V_e(r)$ 
and $T_i(r)$; 
 then, $B_\theta(r)$
can be obtained from  Eq. (\ref{29}), $N_i(r)$ from Eq.
(\ref{25}), $N_e(r)$ from Eq. (\ref{24}) and $T_e(r)$ from Eq. (\ref{25a}).
 Analytic 
solutions,  which are required for determining the portion of 
active particles in equilibria of magnetic confinement systems, 
are constructed in Sec. VI. 
\begin{center}  
{\bf V. PERPENDICULAR NEPs
	 IN EQUILIBRIA OF MAGNETIC CONFINEMENT SYSTEMS}
\end{center}
In this section condition (\ref{13b}) for the existence of perpendicular
NEPs is applied to the equilibria defined in . II and IV.
For  distribution functions  of the form (\ref{19}) 
and the pressure balance relation  (\ref{29}),  the quantity 
$Q_\nu$ [Eq. (\ref{13})] reduces to 
\begin{equation} 
Q_\nu=\frac{N_\nu^\prime}{N_\nu}U_\nu
					  \label{31a}
\end{equation} 
where
\begin{equation} 
U_\nu\equiv 1 - \frac{2}{3} + \eta_\nu\epsilon_{1\nu} +  \epsilon_{2\nu} ,
					  \label{32}
\end{equation} 
\begin{equation} 
\epsilon_{1\nu} \equiv\frac{W_{\nu\perp}}{T_\nu}
	\left(1+\frac{W_{\nu\parallel}}{W_{\nu\perp}}\right),
					  \label{33}
\end{equation} 
\begin{equation} 
\epsilon_{2\nu}\equiv \frac{4\pi}{B^2} \frac{W_{\nu\perp}}{T_\nu} 
		\frac{N_\nu}{N_\nu^\prime} R_\nu,
					  \label{34}
\end{equation} 
and  
\begin{equation} 
R_\nu\equiv P^\prime + \frac{B^2_\theta}{4\pi r}
\left(1+2\frac{W_{\nu\parallel}}{W_{\nu\perp}}\right)
					   \label{35}.
\end{equation}
Depending on the value  of  $T_i/T_e$, the effect of $\bf E$ on
perpendicular NEPs
is examined in the following two regions.
\newline

\noindent
A)\ \ {\em $\frac{\textstyle T_i}{\textstyle T_e}< \beta_c
\approx \frac{\textstyle P}{\textstyle B^2/8\pi}$}
\newline

Assuming the scaling
 (\ref{22}) to hold it can be  shown
that
 the pressure gradient
and a term relating to the curvature of  $B_\theta$
dominate in $Z_\nu$, i.e. 
\begin{equation}
Z_\nu \approx P^\prime + \frac{B_\theta^2}{4\pi r}
\left(1+2\frac{W_{\nu\parallel}}{W_{\nu\perp}}\right)=R_\nu.
					  \label{38}
\end{equation}
Condition (\ref{13b}) can be put in the form 
\begin{equation} 
R_\nu \frac{N_\nu^\prime}{N_\nu}U_\nu<0.
					  \label{38a}
\end{equation}
Relation (\ref{38a}) is identical to the corresponding one
in equilibria with
$\bf E=0$ (Relation (58) of Ref. \cite{ThPf96a}).
 For singly peaked  density and the temperature profiles 
and therefore 
$\eta_\nu>0$ for all $\nu$, which is the most common case in 
equilibria of magnetic confinement systems,
there are two regimes of
NEPs  depending on the sign of $R_\nu$ [Eq. (\ref{35})].
\newline

\noindent
 a)\ \ \underline{{\em $R_\nu<0$}}\ \ \ Condition (\ref{38a}) implies that
$U_\nu<0$ must hold. 
The last two terms of $U_\nu$ [Eqs. (\ref{32})-(\ref{34})]
become non-negative and vanish for
$W_{\nu\parallel}\rightarrow 0$ and $W_{\nu\perp}\rightarrow 0$. 
Consequently,   
 $U_\nu<0$ is satisfied whenever 
\begin{equation}  \eta_\nu >\frac{2}{3}. 
				       \label{39a}
\end{equation} 
  The existence of
perpendicular ion NEPs for any  $k_\perp$ 
 is therefore related to the threshold value of
$2/3$ of the quantity $\eta_\nu$. 
As discussed in
Ref. \cite{ThPf94}, this threshold value is sub-critical in the
sense that it is lower than the critical value
$\eta^c_\nu\approx1$ for linear stability of 
temperature-gradient-driven modes.
\newline

\noindent
b)\ \ \underline{{\em $R_\nu>0$}} \ \ \ The condition for the existence of
perpendicular NEPs becomes $U_\nu>0$. In this case the quantity  
$\eta_\nu\epsilon_{1\nu} +  \epsilon_{2\nu}$ can be either 
positive or negative and therefore no restriction is imposed on $\eta_\nu$.
It may be noted that for plane equilibria it holds that $R_\nu=P^\prime<0$
and therefore the second regime of NEPs is associated with the curvature of
$B_\theta$.
\newline

\noindent

B)\ \ {\em $\frac{\textstyle T_i}{\textstyle T_e} > \beta_c$}
\newline

If the scaling (\ref{22})  holds   
the term $e_\nu \phi^\prime$ related to the 
${\bf E}\times {\bf B}$ drift   
dominates in $Z_\nu$ [Eq. (\ref{12})], i.e. 
\begin{equation} 
Z_\nu\approx-\frac{B^2}{4\pi W_{\nu\perp}} e_\nu\phi^\prime
					   \label{35a}.
\end{equation}
Condition (\ref{13b}) 
becomes then
\begin{equation} 
e_\nu \phi^\prime \frac{N_\nu^\prime}{N_\nu}U_\nu>0.
					  \label{36}
\end{equation} 
Relation (\ref{36}) shows that the existence of perpendicular
NEPs depends on the sign of the
particle species charge and  the polarity of $\bf E$.
Henceforth and up to the subsection ``Reversed-field-pinchlike equilibria"
  $\bf E\neq 0$ will refer to this case ($T_i/T_e>\beta_c$).
 In the edge region 
the  radial electric field is usually negative
\cite{GrBu,FiFu}. 
It is   noted  here 
that the impact of the polarity of an externally induced 
 radial electric field  was
investigated experimentally \cite{WeVa}. It was found that
whereas the energy confinement in H-modes with  $\bf E> 0$ is  at least as 
good as in those with $\bf E< 0$, the ratio of the ion confinement
time to the energy confinement time is about three times lower 
in the former case. We examine therefore in the following
 NEPs for ions and electrons in equilibria with $\phi^\prime>0$. 
\newline

\noindent
\underline{{\em Ions}} \ \ \  In this case $e_\nu \phi^\prime$  is positive 
and condition (\ref{36}) is satisfied whenever  $U_i<0$.  
Depending on the sign of $R_i$,  there are two regimes of
NEPs: a) If $R_i<0$, 
 $U_i<0$ is satisfied whenever $\eta_i>2/3$ and
b) if $R_i>0$ no restriction is imposed on  $\eta_i$. It is pointed out,
however, that for $\bf E = 0$ 
the condition associated with this second regime is $U_i>0$.
As  shown in Sec. VI, this difference affects the fraction of active
 ions.
\newline

\noindent
\underline{\em Electrons} \ \
Condition (\ref{36})
is satisfied whenever $U_e>0$.
 This yields
\begin{equation}
\eta_e < \frac{2}{3}\left(1+\eta_e\epsilon_{1e} + \epsilon_{2e}\right).                                          
					  \label{39b}
\end{equation}
For cold electrons ($W_{e\parallel}\rightarrow 0$ and
$W_{e\perp}\rightarrow 0$) condition (\ref{39b}) implies that 
NEPs exist whenever  $\eta_e<2/3$.  
This indicates that  $\bf E$
has a ``stabilizing" effect on electron NEPs
for large values of
$\eta_e$. Owing to hot electrons however,
 condition (\ref{39b}) does 
not yield  an upper threshold value of $\eta_e$
because   electrons with non-vanishing energy 
 activate NEPs  
 in the regime  where $\eta_e>2/3$. 
 To determine the value of  $\eta_e$ 
 for which half of  the electrons are active,
 condition (\ref{39b}) is written in the form
\begin{equation}
C_e(r)\frac{W_{e\parallel}}{T_e} + D_e(r)\frac{W_{e\perp}}{T_e}> \frac{3}{2}                                  
-\frac{1}{\eta_e},
					  \label{40}
\end{equation}
where 
\begin{equation}
C_e(r)\equiv\left(1 + \frac{2}{r}\frac{B_\theta^2}{B^2}\frac{T_e}{T_e^\prime}\right)
\ \ \ \mbox{and}\ \ \  
D_e(r)\equiv\left[1 + \frac{4\pi}{B^2 }\frac{T_e}{T_e^\prime}
    \left(P^\prime + \frac{B_\theta^2}{4\pi r}\right)\right].
					  \label{41}
\end{equation}
For a magnetic confinement system  it holds that
$B_\theta^2/B^2\approx P/(B^2/8\pi)\equiv \beta$
 with $\max \beta\approx 0.1$
 and therefore  
$ C_e\approx D_e\approx 1$. Consequently,  condition (\ref{40})
implies that
nearly half of the  
electrons are active whenever it holds that 
$3/2 - 1/\eta_e \approx 3/4$.  This yields
\begin{equation}
\eta_e^0 \approx 4/3.
					  \label{43}
\end{equation}
Therefore, if  ${\bf E\neq 0} $  
less than half of the electrons  are active
whenever  $\eta_e>\eta_e^0$ and 
this portion decreases
as $\eta_e$ takes larger values. On the other side, if ${\bf E}=0$
more than half of the electrons are active
whenever $\eta_e>\eta_e^0$ with this  portion increasing as 
$\eta_e$ takes larger values. 


\begin{center} 
{\bf VI.\ \  ANALYTIC  EQUILIBRIUM SOLUTIONS AND ACTIVE PARTICLES}
\end{center}

In this section the portion of   
active particles is determined on the basis of analytic 
 shearless stellaratorlike,  tokamaklike and 
reversed field pinchlike equilibrium solutions. 
\begin{center} 
 {\large \em Shearless stellaratorlike ($\theta$-pinch) equilibria }
\end{center}
We consider the following profiles:
\begin{equation}
P=P(0)(1-\rho^2),
                                                       \label{44a} 
\end{equation}
$N_i=N_i(0)(1-\rho^2)^\xi$ and 
$
T_i=T_i(0)(1-\rho^2)^{1-\xi}
$, where $\rho\equiv r/r_0$ and $r_0$ corresponding to  the plasma surface.
Eqs. (\ref{29}) with $B_\theta\equiv 0$,  $\sum_\nu e_\nu N_\nu = 0$ and 
 $P=\sum_\nu N_\nu T_\nu $ yield then  
\begin{equation}
B_{z} = B_{z0}\left[1 - \beta_0(1-\rho^2)\right]^{1/2},
                                                       \label{44b}
\end{equation}
$$
{\bf j}
 = \nabla B_z \times {\bf e}_z
        = -\frac{B_{z0}}{r_0}\beta_0
           \frac{\rho}{\left[1-\beta_0(1-\rho^2)\right]^{1/2}}\, {\bf e}_\theta,
$$
$N_e=N_e(0)(1-\rho^2)^\xi$ and
$T_e=T_e(0)(1-\rho^2)^{1-\xi}$. Here, $B_{z0}$ is the external constant ``toroidal"
magnetic field ,
$\beta_0\equiv P(0)/(B^2/8\pi)$ and
 the parameter $\xi$ ($0\leq\xi\leq 1$) determines
equilibria with different values of $\eta_\nu$, i.e.,
\begin{equation}
\eta_\nu
\equiv
\frac{\textstyle\partial \ln T_\nu}{\textstyle\partial\ln N_\nu}
=\frac{1-\xi}{\xi}.
                                                     \label{44c}
\end{equation}
Ion and electron NEPs  are now examined separately.
\newline

\noindent
\underline{{\em Ions}} \ \ \ 
Since the magnetic field lines are straight,  it holds that $R_i=P^\prime<0$
and therefore 
ion NEPs  exist only in equilibria     
with  $\eta_i>2/3$.
The pertinent condition
$U_i<0$ becomes 
\begin{equation}
 \frac{W_{i\parallel}}{T_i}
+ \left[1 + \frac{\beta}{2(1-\xi)}(1-\rho^2)\right] 
\frac{W_{i\perp}}{T_i}  < \frac{3}{2} - \frac{1}{\eta_i}.
					      \label{54}
\end{equation}         
Relation (\ref{54}) implies that: 
\begin{enumerate}
\item 
The portion of  active ions
  increases as  $\eta_i$ takes  larger values. 
In particular, for a flat temperature and
 peaked density profile there are no active ions;
 for $\eta_i=1$
one third of the thermal ions are active and  
for $\eta_i=2$  this fraction becomes $2/3$; 
for a flat density and a peaked temperature profile 
($\eta_i\rightarrow \infty)$  
all  ions  are active;
\item The portion of   active ions
 increases from the center $\rho=0$ to the edge region $\rho=1$.
\end{enumerate}  

\noindent
\underline{{\em Electrons}} \ \ \
For $\bf E=0$ the situation is similar to the
foregoing one for ions.  
For $\bf E\neq 0$ the  condition
for the existence of electron NEPs is $U_e>0$ and therefore
the  fractions  of  
active electrons and ions 
are complementary to each other.
Thus, 
as also discussed in Sec. V, 
the electric field  stabilizes 
electron NEPs
for   $\eta_e> 4/3$,   
e.g.,   
for the   
equilibrium profiles (\ref{44a})-(\ref{44c})
$1/3$ of the thermal electrons are active
when $\eta_e=2$,  while
the  corresponding fraction for the  equilibrium with ${\bf E=0}$ is $2/3$.
In addition, the fraction of active electrons
{\em decreases} from the center to the edge. This
indicates that in the presence of $\bf E$ self-sustained turbulence
associated with electron NEPs should be reduced in the edge.
\newpage 

\begin{center} 
 {\large \em  Tokamaklike (screw pinch) equilibria}
\end{center}
 
The following  profiles correspond to a special solution of Eq. (\ref{29}):
\begin{equation} 
B_z= \left[B^2_z(0) +8\pi P(0)(1-\alpha^2)\rho^2\right]^{1/2};
					   \label{57}
\end{equation} 
 $\alpha$  is a parameter which describes the magnetic properties 
 of the plasma, i.e. the plasma is diamagnetic for  $\alpha<1$ and
paramagnetic for $\alpha>1$;
\begin{equation} 
B_\theta = 2 \sqrt{\pi P(0)} \alpha \rho;
					  \label{58}
\end{equation} 
constant axial current density; $N_\nu=N_\nu(0)(1-\rho^2)^\xi$ and
 $T_\nu=T_\nu(0)(1-\rho^2)^{1-\xi}$ with $\nu=i, e$. 
 
Ion and electrons NEPs are  now examined for $\eta_\nu=1$, which is close to linear stability
threshold for gradient temperature driven modes. 
\newline

\noindent
\underline{{\em Ions}} \ \ \ 
For  $\bf E=0$  
the portion of   active ions
is determined by conditions
(\ref{14}) and (\ref{15}) which
respectively become
\begin{equation} 
\frac{W_{i\parallel}}{W_{i\perp}} <\frac{1}{2}
  \left(\frac{2}{\alpha^2}-1\right)
 \ \ \mbox{and} \ \ 
 \left[1-\frac{1}{2}\beta(1-\rho^2)(\alpha^2-2)
 \right]\frac{W_{i\perp}}{T_i} +
 \left[1-\beta\alpha^2(1-\rho^2)
 \right]\frac{W_{i\parallel}}{T_i}  < \frac{1}{2} 
					      \label{62}
\end{equation} 
and
\begin{equation} 
\frac{W_{i\parallel}}{W_{i\perp}} >\frac{1}{2}
  \left(\frac{2}{\alpha^2}-1\right)
 \ \ \mbox{and} \ \ 
 \left[1-\frac{1}{2}\beta(1-\rho^2)(\alpha^2-2)
 \right]\frac{W_{i\perp}}{T_i} +
 \left[1-\beta\alpha^2(1-\rho^2)
 \right]\frac{W_{i\parallel}}{T_i}  > \frac{1}{2}. 
					      \label{63}
\end{equation} 
 Relations (\ref{62}) and (\ref{63}) imply that:
\begin{enumerate}
\item The portion of active ions increases with $\alpha$, i.e.,
it is smaller in a 
 paramagnetic and larger in a  diamagnetic  system. 
The particular  cases of a strongly diamagnetic plasma  
($\alpha\rightarrow 0$),
of an equilibrium with constant ``toroidal" magnetic field  
($\alpha^2=1$) and of a paramagnetic plasma ($\alpha^2=2$) are illustrated,
in Figs. 1, 2 and 3, respectively.  The fractions of active
 ions are nearly $1/3$ for $\alpha\rightarrow 0$, $1/2$ 
for $\alpha^2=1$ and $2/3$ for $\alpha^2=2$.  It is noted
that for $\alpha\rightarrow 0$ only the branch (\ref{62}), 
associated with the threshold value $\eta_i=2/3$ contributes
 while for $\alpha^2=2$ exclusively the branch 
(\ref{63})   associated with the curvature of the poloidal field lines
contributes.
\vspace{1.0cm}
\begin{center}
\begin{picture}(7,7)
\put(0,1){\line(0,1){6}}
\put(0,1){\line(1,0){5}}
%
%
\put(3.5,1){\line(-1,1){3.5}}
\put(3.5,1){\line(-2,3){3.5}}
\put(-0.5,0.5){\large (0,0)}
\put(-1,7){\Large$\frac{W_{i\perp}}{T_e}$}
\put(5.0,0.2){\Large$\frac{W_{i\parallel}}{T_i}$}
\put(3.7,1.5){\Large$\frac{1}{2}$}
\put(0.5,6.0){\Large$\frac{1}{2a_0(1)}$}
\put(-1.5,4.5){\Large$\frac{1}{2a_0(0)}$}
\put(3.0,4.5){\fbox{\Large$\alpha\  \rightarrow\  0$}}
\put(3.0,5.5){\fbox{\Large${\bf E=0}$}}
%
\multiput(.2,1.2)(.3,0){10}{\circle*{.1}}
\multiput(.2,1.5)(.3,0){9}{\circle*{.1}}
\multiput(.2,1.8)(.3,0){8}{\circle*{.1}}
\multiput(.2,2.1)(.3,0){7}{\circle*{.1}}
\multiput(.2,2.4)(.3,0){6}{\circle*{.1}}
\multiput(.2,2.7)(.3,0){5}{\circle*{.1}}
\multiput(.2,3.0)(.3,0){4}{\circle*{.1}}
\multiput(.2,3.3)(.3,0){3}{\circle*{.1}}
\multiput(.2,3.3)(.3,0){2}{\circle*{.1}}
\multiput(.2,3.6)(.3,0){2}{\circle*{.1}}
\multiput(.2,3.9)(.3,0){1}{\circle*{.1}}
%
%
\multiput(2.9,1.8)(.3,0){1}{\circle{.1}}
\multiput(2.6,2.1)(.3,0){1}{\circle{.1}}
\multiput(2.4,2.4)(.3,0){1}{\circle{.1}}
\multiput(2.2,2.7)(.3,0){1}{\circle{.1}}
\multiput(1.7,3.0)(.3,0){2}{\circle{.1}}
\multiput(1.5,3.3)(.3,0){2}{\circle{.1}}
\multiput(1.0,3.6)(.3,0){3}{\circle{.1}}
\multiput(0.8,3.9)(.3,0){3}{\circle{.1}}
\multiput(0.5,4.2)(.3,0){3}{\circle{.1}}
\multiput(0.2,4.5)(.3,0){3}{\circle{.1}}
\multiput(0.2,4.8)(.3,0){3}{\circle{.1}}
\multiput(0.2,5.1)(.3,0){2}{\circle{.1}}
\multiput(0.2,5.4)(.3,0){2}{\circle{.1}}
\multiput(0.2,5.7)(.3,0){1}{\circle{.1}}
\end{picture}
\end{center}

\noindent
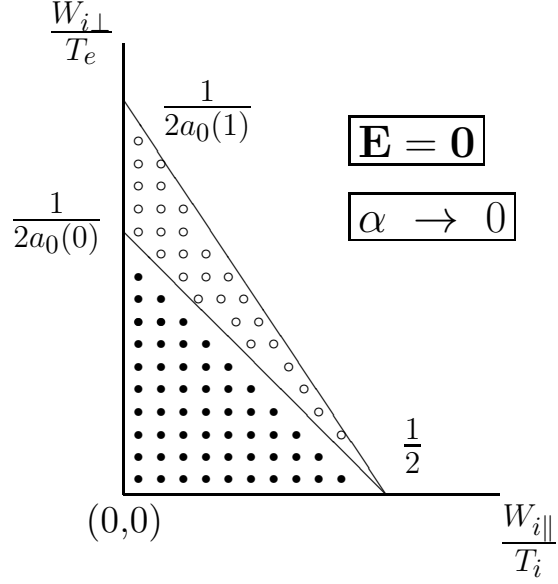
FIG. 1. The portion of   active ions
       for a strongly diamagnetic plasma with ${\bf E=0}$
       and $\eta_i=1$ which is deduced from 
       Eq. (\ref{62}) [$\alpha_0(\rho)\equiv 1 + \beta(1-\rho^2)$].
       The dotted area stands for the active particles at the 
       center ($\rho=0$), while the area filled
      by circles for  the  additional  active particles 
      at the edge ($\rho=1$). 
\vspace{1cm} 
\item In all regimes the fraction of active ions 
increases from  
the center to the edge. In Figs. 1, 2 and 3 
the dotted area stands for the active particles at the 
center ($\rho=0$), while the area filled
by circles for  the  additional active particles 
at the edge ($\rho=1$).  
\end{enumerate}
It is noted here that for $\bf E=0$ similar results hold for electrons.
\vspace{1cm}
\begin{center}
\begin{picture}(7,7)
\put(0,1){\line(0,1){6}}
\multiput(3.2,3,8)(.3,0){2}{\circle*{.1}}
%
\multiput(3.0,1.3)(.3,0){1}{\circle{.1}}
\multiput(2.7,1.6)(.3,0){1}{\circle{.1}}
\multiput(2.5,1.9)(.3,0){1}{\circle{.1}}
\multiput(2.2,2.2)(.3,0){1}{\circle{.1}}
%
\multiput(1.4,3.2)(.3,0){1}{\circle{.1}}
\multiput(1.2,3.5)(.3,0){1}{\circle{.1}}
\multiput(1.0,3.8)(.3,0){1}{\circle{.1}}
\multiput(0.7,4.1)(.3,0){1}{\circle{.1}}
\multiput(0.3,4.4)(.3,0){2}{\circle{.1}}
\multiput(0.3,4.7)(.3,0){1}{\circle{.1}}
\multiput(0.1,5.0)(.3,0){1}{\circle{.1}}
%
\put(0,1){\line(1,0){6}}
\put(3.5,1){\line(-1,1){3.5}}
\put(2.8,1){\line(-3,5){2.8}}
\put(0,1){\line(1,1){3.}}
\put(-0.5,0.5){\large(0,0)}
\put(-1,7.0){\Large$\frac{W_{i\perp}}{T_i}$}
\put(6.0,0){\Large$\frac{W_{i\parallel}}{T_i}$}
\put(3.5,1.5){\Large$\frac{1}{2b_1(0)}$}
\put(1.8,0.3){\Large$\frac{1}{2b_1(1)}$}
\put(.3,6){\Large$\frac{1}{2a_1(1)}$}
\put(-1.5,4.5){\Large$\frac{1}{2a_1(0)}$}
\put(2.0,4.5){$W_{i\parallel}=\frac{1}{2}W_{i\perp}$}
\put(4.2,5.8){\fbox{\Large$\alpha^2=1$}}
\put(4.2,6.8){\fbox{\Large${\bf E=0}$}} 
%
\multiput(.2,1.5)(.3,0){1}{\circle*{.1}}
\multiput(.2,1.7)(.3,0){2}{\circle*{.1}}
\multiput(.2,2.0)(.3,0){3}{\circle*{.1}}
\multiput(.2,2.3)(.3,0){4}{\circle*{.1}}
\multiput(.2,2.6)(.3,0){5}{\circle*{.1}}
\multiput(.2,2.9)(.3,0){5}{\circle*{.1}}
\multiput(.2,3.2)(.3,0){4}{\circle*{.1}}
\multiput(.2,3.5)(.3,0){3}{\circle*{.1}}
\multiput(.2,3.8)(.3,0){2}{\circle*{.1}}
\multiput(.2,4.1)(.3,0){1}{\circle*{.1}}
%
\multiput(3.2,1.7)(.3,0){1}{\circle*{.1}}
\multiput(2.9,2.0)(.3,0){3}{\circle*{.1}}
\multiput(2.6,2.3)(.3,0){5}{\circle*{.1}}
\multiput(2.3,2.6)(.3,0){6}{\circle*{.1}}
\multiput(2.5,2.9)(.3,0){6}{\circle*{.1}}
\multiput(2.7,3.2)(.3,0){5}{\circle*{.1}}
\multiput(3.0,3,5)(.3,0){3}{\circle*{.1}}
\end{picture}
\end{center}
\vspace{0.5cm}

\noindent
FIG. 2. The portion of   active
        ions for the equilibrium with ${\bf E=0}$, $\eta_i=1$
        and $B_z=$ constant, which is deduced from Eqs. (\ref{62})
        and (\ref{63})
           [$\alpha_1(\rho) \equiv 1 +(\beta/2) (1-\rho^2),
                       \  b_1(\rho) \equiv  1 -\beta(1-\rho^2)$].
\vspace{0.5cm}
\begin{center}
\begin{picture}(7,7)
\put(0,1){\line(0,1){5}}
\put(0,1){\line(1,0){7}}
\put(0,4.5){\line(1,-1){3.5}}
\put(0,4.5){\line(3,-2){5.2}}
\put(-0.5,0.5){\large (0,0)}
\put(-1,6){\Large$\frac{W_{i\perp}}{T_i}$}
\put(7.0,0.2){\Large$\frac{W_{i\parallel}}{T_i}$}
\put(-0.7,4.5){\Large$\frac{1}{2}$}
\put(2.5,0.2){\Large$\frac{1}{2b_2(1)}$}
\put(4.7,2.0){\Large$\frac{1}{2b_2(0)}$}
\put(3.0,4.5){\fbox{\Large$\alpha^2 =2$}}
\put(3.0,5.5){\fbox{\Large$\bf E=\bf 0$}}  
%
\multiput(5.2,1.2)(.3,0){4}{\circle*{.1}}
\multiput(4.7,1.5)(.3,0){4}{\circle*{.1}}
\multiput(4.4,1.8)(.3,0){1}{\circle*{.1}}
\multiput(4.0,2.1)(.3,0){1}{\circle*{.1}}
\multiput(3.6,2.4)(.3,0){3}{\circle*{.1}}
\multiput(3.1,2.7)(.3,0){4}{\circle*{.1}}
\multiput(2.6,3.0)(.3,0){5}{\circle*{.1}}
\multiput(2.0,3.3)(.3,0){6}{\circle*{.1}}
\multiput(1.5,3.6)(.3,0){8}{\circle*{.1}}
\multiput(1.2,3.9)(.3,0){5}{\circle*{.1}}
\multiput(0.7,4.2)(.3,0){6}{\circle*{.1}}
\multiput(0.2,4.5)(.3,0){7}{\circle*{.1}}
\multiput(0.2,4.8)(.3,0){7}{\circle*{.1}}
\multiput(0.2,5.1)(.3,0){7}{\circle*{.1}}
\multiput(0.2,5.4)(.3,0){8}{\circle*{.1}}
%
%
\multiput(3.5,1.2)(.3,0){5}{\circle{.1}}
\multiput(3.1,1.5)(.3,0){4}{\circle{.1}}
\multiput(2.9,1.8)(.3,0){3}{\circle{.1}}
\multiput(2.6,2.1)(.3,0){3}{\circle{.1}}
\multiput(2.3,2.4)(.3,0){2}{\circle{.1}}
\multiput(2.0,2.7)(.3,0){2}{\circle{.1}}
\multiput(1.9,3.0)(.3,0){1}{\circle{.1}}
\multiput(1.5,3.3)(.3,0){1}{\circle{.1}}
\multiput(1.0,3.6)(.3,0){1}{\circle{.1}}
\end{picture}
\end{center}
\vspace{0.5cm}
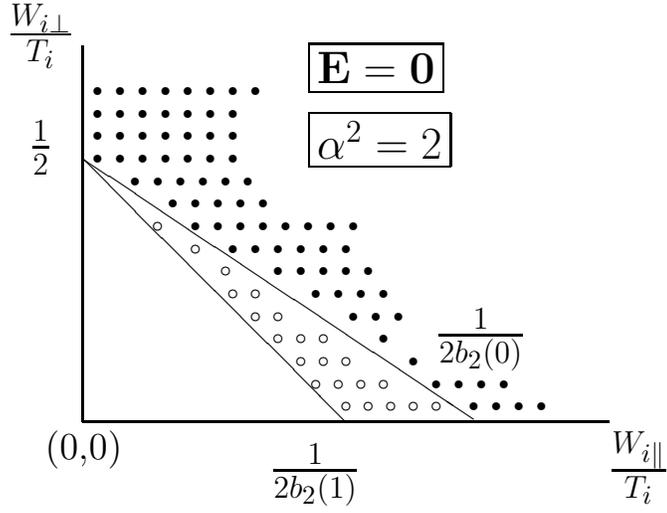
FIG. 3. The portion of   active 
	ions for the equilibrium of a paramagnetic plasma 
	with ${\bf E}={\bf 0}$ and $\eta_e =1$,
	which is deduced from Eq. (\ref{63}) 
	[$b_2(\rho) \equiv 1 -2\beta(1-\rho^2)$]. 

\vspace{1.5cm}
 
For $\bf E\neq 0$, active ions  
obtain from condition $U_i< 0$ 
(irrespective of  
the sign of $R_i$) 
which leads to 
\begin{equation}
 \left[1-\frac{1}{2}\beta(1-\rho^2)(\alpha^2-2)
 \right]\frac{W_{i\perp}}{T_i} +
 \left[1-\beta\alpha^2(1-\rho^2)
 \right]\frac{W_{i\parallel}}{T_i}  < \frac{1}{2} .
					      \label{64}
\end{equation} 
\vspace{0.5cm}
\begin{center}
\begin{picture}(7,7)
\put(0,1){\line(0,1){6}}
\put(0,1){\line(1,0){6}}
\put(3.5,1){\line(-1,1){3.5}}
\put(2.8,1){\line(-3,5){2.8}}
\put(-0.5,0.5){\large(0,0)}
\put(-1,7.0){\Large$\frac{W_{i\perp}}{T_i}$}
\put(6.0,0){\Large$\frac{W_{i\parallel}}{T_i}$}
\put(3.5,1.5){\Large$\frac{1}{2b_4(0)}$}
\put(1.8,0.3){\Large$\frac{1}{2b_4(1)}$}
\put(.3,6){\Large$\frac{1}{2a_4(1)}$}
\put(-1.5,4.5){\Large$\frac{1}{2a_4(0)}$}
\put(4.2,5.8){\fbox{\Large$\alpha^2=1$}}
\put(4.2,6.8){\fbox{\Large$\bf E\neq 0$}} 
%
\multiput(.2,1.4)(.3,0){8}{\circle*{.1}}
\multiput(.2,1.7)(.3,0){7}{\circle*{.1}}
\multiput(.2,2.0)(.3,0){7}{\circle*{.1}}
\multiput(.2,2.3)(.3,0){6}{\circle*{.1}}
\multiput(.2,2.6)(.3,0){6}{\circle*{.1}}
\multiput(.2,2.9)(.3,0){5}{\circle*{.1}}
\multiput(.2,3.2)(.3,0){4}{\circle*{.1}}
\multiput(.2,3.5)(.3,0){3}{\circle*{.1}}
\multiput(.2,3.8)(.3,0){2}{\circle*{.1}}
\multiput(.2,4.1)(.3,0){1}{\circle*{.1}}
%
%
\multiput(2.9,1.3)(.3,0){1}{$\star$}
\multiput(2.6,1.6)(.3,0){1}{$\star$}
\multiput(2.3,1.9)(.3,0){1}{$\star$} 
\multiput(2.1,2.1)(.3,0){1}{$\star$}
%
\multiput(1.4,3.2)(.3,0){1}{\circle{.1}}
\multiput(1.2,3.5)(.3,0){1}{\circle{.1}}
\multiput(1.0,3.8)(.3,0){1}{\circle{.1}}
\multiput(0.7,4.1)(.3,0){1}{\circle{.1}}
\multiput(0.3,4.4)(.3,0){2}{\circle{.1}}
\multiput(0.3,4.7)(.3,0){1}{\circle{.1}}
\multiput(0.1,5.0)(.3,0){1}{\circle{.1}}
\end{picture}
\end{center}
\vspace{0.5cm}
FIG. 4. The portion of   active
	ions for the equilibrium with ${\bf E\neq 0}$, $\eta_i=1$
	and $B_z=$ constant, which is deduced from Eq. (\ref{64})
	[$\alpha_4(\rho) \equiv 1 +(\beta/2) (1-\rho^2),
		       \  b_4(\rho) = 1 -\beta(1-\rho^2)$].
	The excess portion at the edge indicated by circles nearly
	compensates for the excess portion at the center indicated by
	stars.
\vspace{0.5cm}

\noindent
Relation (\ref{64}) implies  that: 
\begin{enumerate}
\item The portion of active ions is nearly independent
      of the magnetic properties of the plasma; it is approximately 
      $1/3$  for any value of $\alpha$.
\item 
      The portion of active ions
      can either be nearly independent of $\rho$, e.g.,   
      for an equilibrium
      with  constant $B_z$ ($\alpha^2=1$) (Fig. 4)
      or decreases from the center to the edge, e.g. 
      for a paramagnetic plasma $\alpha^2=2$ (Fig. 5), while this
      portion always increases for equilibria with $\bf E=0$.  
\end{enumerate}
Thus, 
$\bf E$ leads to a reduction of active  ions.
\vspace{1cm} 
\begin{center}
\begin{picture}(7,7)
\put(0,1){\line(0,1){5}}
\put(0,1){\line(1,0){7}}
\put(0,4.5){\line(1,-1){3.5}}
\put(0,4.5){\line(3,-2){5.2}}
\put(-0.5,0.5){\large (0,0)}
\put(-1,6){\Large$\frac{W_{i\perp}}{T_i}$}
\put(7.0,0.2){\Large$\frac{W_{i\parallel}}{T_i}$}
\put(-0.7,4.5){\Large$\frac{1}{2}$}
\put(2.5,0.2){\Large$\frac{1}{2b_5(1)}$}
\put(4.7,2.0){\Large$\frac{1}{2b_5(0)}$}
\put(3.0,4.5){\fbox{\Large$\alpha^2 =2$}}
\put(3.0,5.5){\fbox{\Large{$\bf E\neq 0$}}}  
%
\multiput(.2,1.2)(.3,0){10}{\circle*{.1}}
\multiput(.2,1.5)(.3,0){9}{\circle*{.1}}
\multiput(.2,1.8)(.3,0){8}{\circle*{.1}}
\multiput(.2,2.1)(.3,0){7}{\circle*{.1}}
\multiput(.2,2.4)(.3,0){6}{\circle*{.1}}
\multiput(.2,2.7)(.3,0){5}{\circle*{.1}}
\multiput(.2,3.0)(.3,0){4}{\circle*{.1}}
\multiput(.2,3.3)(.3,0){3}{\circle*{.1}}
\multiput(.2,3.3)(.3,0){2}{\circle*{.1}}
\multiput(.2,3.6)(.3,0){2}{\circle*{.1}}
\multiput(.2,3.9)(.3,0){1}{\circle*{.1}}
%
\multiput(3.2,1.2)(.3,0){4}{$\star$}
\multiput(3.1,1.5)(.3,0){4}{$\star$}
\multiput(2.9,1.8)(.3,0){3}{$\star$}
\multiput(2.6,2.1)(.3,0){3}{$\star$}
\multiput(2.3,2.4)(.3,0){2}{$\star$}
\multiput(2.0,2.7)(.3,0){2}{$\star$}
\multiput(1.7,3.0)(.3,0){1}{$\star$}
\multiput(1.4,3.3)(.3,0){1}{$\star$}
\multiput(0.9,3.6)(.3,0){1}{$\star$}
\end{picture}
\end{center}
\vspace{0.5cm}

\noindent
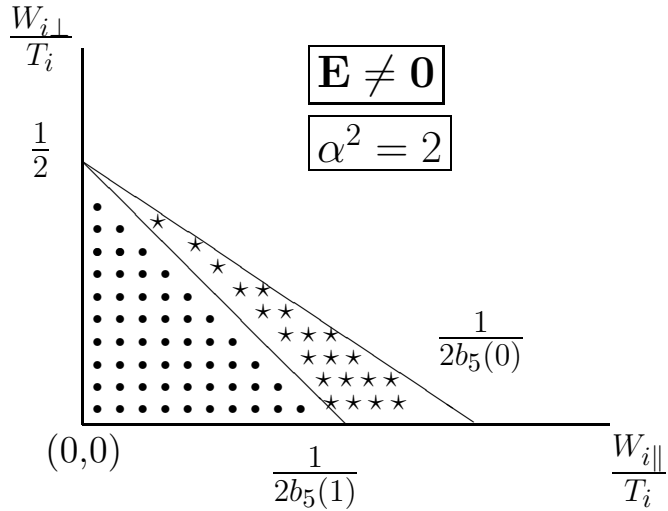
FIG. 5. The portion of   active ions
       for a strongly diamagnetic plasma with ${\bf E\neq 0}$
       and $\eta_i=1$
       deduced from Eq. (\ref{64}) 
       [$\alpha_5(\rho)=1 -2\beta(1-\rho^2)$]. The area filled 
       by stars represents the excess portion at the plasma  
       center.

\vspace{1.0cm}

\noindent
\underline{{\em Electrons}} \ \ \ 
Recalling that the portion  of 
active electrons is the same  as that  of active ions  when $\bf E=0$,
and complementary when $\bf E\neq 0$, respectively, 
the  former  portion can  be determined on the basis of
the foregoing analysis for ions. Thus, in addition to the
stabilizing effect of $\bf E$ for  $\eta_e>4/3$,  
the fraction  of active electrons 
(a) becomes nearly independent of the 
magnetic properties of the plasma and 
(b) can decrease from the center to the 
edge, e.g., for the most common
case of a diamagnetic plasma.
\begin{center} 
 {\large \em Reversed-field-pinchlike (force-free) equilibria}
\end{center}

The solution of Eq. (\ref{29}) with 
 $P^\prime = 0$ leads to   
$B_{z}=B_{z}(0)J_{0}(\rho)$ and $B_{\theta}=B_{z}(0)J_{1}(\rho)$,
where $J_0$ and $J_1$ are Bessel functions.  
These profiles   satisfactorily describe the central region of the relaxed
state of a reversed-field pinch \cite{Ta}.  
By appropriately assigning $V_i(r)-V_e(r)$, 
one can derive  equilibria  with a variety 
of density  and temperature
profiles for which NEPs
 exist and a considerable fraction of active  ions and
electrons are involved. From the equilibria considered it turns out
that $\bf E$ (a) does not affect the electron NEPs and
(b) enhances the fraction of active ions. 

As an example, we consider the most common  equilibrium
with  constant density and temperature profiles:
\begin{equation} 
N_\nu=N_{\nu 0},\ \ \ \   T_\nu=T_{\nu 0}.
\end{equation} 
For $\bf E=0$, with the aid  of relation (\ref{38})
 condition (\ref{13b})  becomes 
\begin{equation} 
\frac{W_{\nu\perp}}{T_{\nu 0}}
\frac{B_\theta^2}{\rho B^2}
\left(1 + 2\frac{W_{\nu\parallel}}{W_{\nu\perp}}\right)<0
					    \label{65}
\end{equation} 
for any particle species $\nu$.
Therefore there are neither ion nor electron NEPs.

If  $\bf E\neq 0$, NEPs
exist whenever the condition
\begin{equation} 
\frac{e_\nu \phi^\prime}{T_{\nu 0}} \frac{B_\theta^2}{\rho B^2}
\left(1 + 2\frac{W_{\nu\parallel}}{W_{\nu\perp}}\right)>0,
					    \label{66}
\end{equation} 
following from relations (\ref{13b}) and (\ref{35a}), is satisfied.
Owing to the presence of the particle species charge in condition ({\ref{66}),
for $\phi^\prime>0$ all ions are active,   
while the active electrons are  not affected.
\begin{center}
{\bf VII.\ \ CONCLUSIONS}
\end{center}
The impact of a radial electric field 
on negative-energy perturbations (NEPs) in 
cylindrical
equilibria of  magnetically confined plasmas was investigated
within the framework of linearized dissipationless  Maxwell-drift 
kinetic theory. The investigation consisted in evaluating the  
general expression for the 
second-order perturbation energy derived by Pfirsch and Morrison   
for the equilibria under consideration and   
 for vanishing initial field perturbations;  then the  conditions
for the existence  of NEPs were obtained. 

The electric field 
$\bf E$ does not affect the  following condition for perturbations 
with wave vectors parallel and oblique to the equilibrium magnetic field 
($k_\parallel\neq 0$): 
If the equilibrium guiding center distribution function
$f_{g\nu}^{(0)}(r, v_\parallel, \mu)$ of at least one particle species $\nu$
 satisfies the relation  $v_\parallel 
(\partial f^{(0)}_{g\nu}/ \partial v_\parallel) >0$
locally in $r, v_\parallel$ and $\mu$, parallel and oblique NEPs  
exist with no
essential restriction on  ${\bf k}$. The condition for 
the existence of perpendicular NEPs ($k_\parallel = 0$), which holds  regardless of the sign of 
$v_\parallel(\partial f_{g\nu}^{(0)}/\partial v_\parallel)$}, 
  is modified.  For  
  $|e_i\phi|\approx T_i$
the effect  of $\bf E$ on perpendicular NEPs depends on the 
value  of $T_i/T_e $,  i.e.,
 a) for $T_i/T_e < \beta_c\approx P/(B^2/8\pi)$ the electric field  has no   effect,
 and b)
for $T_i/T_e> \beta_c$, a case which is of operational interest in magnetic confinement 
systems,   the existence of perpendicular NEPs depends on
the sign of the particle species charge and 
  the  polarity of $\bf E$ [Relation (\ref{13b})]. 
 For $\bf E < 0$ it was found that: 
\begin{enumerate}
\item For cylindrical tokamaklike  equilibria 
described by local shifted Maxwellian distribution functions 
and singly peaked pressure  profiles  
 there exist 
 two regimes of NEPs  for both ions
and electrons. 
One  regime is associated with  the
curvature of the poloidal magnetic field. 
In the other regime the threshold  value $2/3$ of
$\eta_i\equiv \frac{\textstyle\partial \ln T_i}{\textstyle \partial \ln N_i}$
is involved for ion NEPs,  as in equilibria with $\bf E=0$,
while a critical value of $\eta_e$ does not occur for the existence
of electron NEPs. 
However, $\bf E$ has the following ``stabilizing"
effects on both particle species:  
\begin{itemize}
\item The portion of
   particles associated with NEPs (active particles)
is nearly independent of the plasma  magnetic properties,  i.e. it is
nearly the same in a diamagnetic  and  in a paramagnetic plasma,
while in equilibria with $\bf E=0$ this portion is much larger in a 
paramagnetic  than in a diamagnetic plasma. 
\item  The portion of active particles  
      can be either constant or decreases from the center
      to the edge, e.g., in the case of active electrons of a  
      diamagnetic plasma, while it always increases 
      in the corresponding   equilibria with $\bf E=0$.
\end{itemize}      
In particular,  the fraction of  active electrons decreases with increasing
$\eta_e$ and for $\eta_e>\eta_0\approx4/3$ the electric field 
stabilizes  electron NEPs in the sense that the fraction 
of active electrons becomes smaller than the one corresponding to equilibria 
with ${\bf E=0}$. 
\item In  shearless stellaratorlike  equilibria 
described by local Maxwellian distribution functions and  pressure profiles 
identical to those of tokamaklike-equilibria,  $\bf E$ leads to   similar
stabilizing effects on electron NEPs; namely, it
reduces the fraction of active electrons (a)
for $\eta_e>\eta_0\approx 4/3$ and (b) from the center to the edge.
\end{enumerate}
In addition, irrespective of the value of $T_i/T_e$, ${\bf E}$ 
      does not affect the electron NEPs in reversed-field pinchlike equilibria
      but  ``destabilizes" the  ion NEPs in the sense that it enhances  the 
      portion of 
       active ions.  For example,  
       for an equilibrium with constant density and temperature
      profiles all  ions are active in the presence of ${\bf E}$, 
      while there are not
      active ions  when  $\bf E = 0$.

The present results indicate that a radial electric field leads to a reduction
 of the NEPs activity in the edge region of tokamaks and stellarators.  
For electrons,  which 
may mainly contribute to  anomalous transport, 
this reduction is more pronounced.

Finally, it may be noted that
according to the results of our previous \cite{ThPf96a,ThPf96b} 
 and in the present studies,   
 the curvature of the poloidal  magnetic field is unfavorable
 in the sense that it gives rise to an increase
of   NEPs activity.
It can be speculated that this is true for an arbitrary magnetic field
configuration.
To check this conjecture, it is interesting to 
investigate NEPs in a toroidal  equilibrium, e.g.,  a tokamak,
in   which the toroidal magnetic field   is favorably curved  on the inside
and unfavorably  on the outside of the torus.
Such a study  
might also reveal the effect of toroidicity on other aspects
of  NEPs, e.g., the threshold 
value $\eta_\nu=2/3$. 
\begin{center} 
{\Large \bf Acknowledgments}
\end{center}
 
G.N.T. would like to thank D. Correa   for
useful discussions and for a critical reading of the manuscript,
and  H. Tasso and  H. Weitzner for useful discussions. 
Part  of the work was conducted during a visit of G.N.T.
to the Max-Planck-Institut f\"ur
Plasmaphysik, Garching. The hospitality  provided by  this Institute is
appreciated.

The same author acknowledges support by EURATOM through the fixed 
contribution contract ERB 5004 CT 96 0029.


\begin{thebibliography}{99}
\bibitem{BrMo} A. J. Brizard, J. J. Morehead and A. N. Kaufman, 
	       Phys. Rev. Lett. {\bf 77}, 1500 (1996).
\bibitem{We}J. Weiland and H. Wilhelmson, {\it Coherent Non-linear Interaction
	       of  Waves in Plasmas} (Pergamon, New York, 1977).
\bibitem{Wi}H. Wilhelmson, Nucl. Phys. A {\bf 518}, 84 (1990).
\bibitem{Ko}M. Kotschenreuther {\it et al., in Plasma Physics and Controlled
Nuclear Fusion Research 1986} (International Atomic Energy Agency, Vienna,
1987), Vol.  2, p. 149.
\bibitem{MoPf89}P. J. Morrison and D. Pfirsch, 
		Phys. Rev. A {\bf 40}, 3898 (1989).
\bibitem{MoPf90}P. J. Morrison and D. Pfirsch, Phys. Fluids B {\bf 2}, 
	       1105 (1990).
\bibitem{PfMo} D. Pfirsch and P. J. Morrison, Phys.\ Fluids B {\bf 3},
		271, 1991.
\bibitem{CoPf92} D. Correa-Restrepo and D. Pfirsch, Phys. Rev. A {\bf 45},
		   2512 (1992).
\bibitem{CoPf93} D. Correa-Restrepo and D. Pfirsch, Phys. Rev. E {\bf 47},
		   545 (1993).
\bibitem{CoPf94a} D. Correa-Restrepo and D. Pfirsch, Phys. Rev. E {\bf 49},
		   692 (1994).
\bibitem{CoPf94b} D. Correa-Restrepo and D. Pfirsch, 
		  In {\it Proc. of the 21st EPS Conf. Control. Fusion and Plasma
		  Phys., Montpellier, 1994}, edited by  E. Joffrin,
		  P. Platz, P. E. Stott. ECA 18B, (European Physical 
		  Society, Geneva 1398, 1994) p.1398.
\bibitem{CoPf97} D. Correa-Restrepo and D. Pfirsch, Phys. Rev. E {\bf 55},
                   7449 (1997).
 \bibitem{NoPaWe}   H. Nordman, V.\ P.\ Pavlenko, and J. Weiland,
		 Phys.\ Fluids B {\bf 5}, 402 (1993).
\bibitem{CoPf93b}  D. Pfirsch and D. Correa-Restrepo, Phys. Rev. E {\bf 47},
		   1947 (1993).
\bibitem{Pf}      D. Pfirsch, Phys. Rev. E {\bf 48},
		   1428 (1993).
\bibitem{PfWe}  D. Pfirsch and H. Weitzner, Phys. Rev. E {\bf 49},
		   3368 (1994).
\bibitem{ThPf94} G. N. Throumoulopoulos and D. Pfirsch, Phys. Rev. E {\bf 49}, 
	       3290 (1994).
\bibitem{ThPf96a} G. N. Throumoulopoulos and D. Pfirsch, Phys. Rev. E {\bf 53}, 
	       2767, (1996).
\bibitem{ThPf96b} G. N. Throumoulopoulos and D. Pfirsch, Negative-energy
		 perturbations in circularly cylindrical equilibria
		 within the framework of Maxwell-drift kinetic theory,
		 Technical Report 6/337, Max-Planck-Institut f\"ur 
		 Plasmaphysik, Garching bei M\"unchen, Germany, 1996.
\bibitem{Un}   M. Unverzagt, Bedeutung der Energie dynamisch zug\"anglicher 
	       linearer St\"orungen eines Gleichgewichts f\"ur dessen 
	       Stabilit\"at angewandt auf quasineutrale electrostatische oder 
		magnetodynamische Driftst\"orungen, PhD Thesis,
		Max-Planck-Institut f\"ur Plasmaphysik, Garching (1996).
\bibitem{Li} R. G. Littlejohn, J. Plasma Phys.\ {\bf 29}, 111 (1983).
\bibitem{Wim} H. K. Wimmel, Z. Naturforsch. {\bf 38a}, 601 (1983).
\bibitem{Di}  P. A. M. Dirac, Can. J. Math. {\bf 2}, 129 (1950); 
	      Proc. R. Soc. London Ser. A
	      {\bf 246}, 326 (1958); K. Sundermeyer, {\it Constraint Dynamics},
	       Lecture Notes in Physics, edited by H. Araki, 
	       J. Ehlers,. K. Hepp, 
	       R. Kippenhahn, H. A. Weidenm\"uller, 
	       and J. Zittartz (Springer, Berlin,
	       1982), Vol. 169.
\bibitem{CoPfWi} D. Correa-Restrepo, D. Pfirsch and H. K. Wimmel,
		  Physica {\bf 136A}, 453 (1986).
\bibitem{PfCo}  D. Pfirsch and D. Correa-Restrepo,
		  Plasma Phys. Control. Fusion {\bf 38}, 71 (1996).  
\bibitem{GrBu} R. J. Groebner, K. H. Burrell, and R. P. Seraydarian,  
	       Phys. Rev. Lett. {\bf 64}, 3015 (1990).
\bibitem{FiFu} A. R. Field, G. Fussman, J. V. Hofmann and 
	       the ASDEX Team, Nucl. Fusion  
	       {\bf 32}, 1191 (1992).
\bibitem{WeVa} R. R. Weynants {\it et. al}, Nucl. Fusion {\bf 32}, 837 (1992).
\bibitem{Ta} J. B. Taylor, Rev. Mod. Phys. {\bf 58}, 741 (1986).
%
\end{thebibliography}
\end{document}